\documentclass[aps,pra,twocolumn,superscriptaddress,amsmath,amssymb]{revtex4-2}
\usepackage{graphicx}
\usepackage{physics}
\usepackage{lipsum}
\usepackage{hyperref}
\usepackage{url}
\usepackage{amsmath}
\usepackage{amsfonts}
\usepackage{amssymb}
\usepackage{siunitx}
\usepackage[capitalize]{cleveref}
\begin{document}
\title{Cavity Characterization for the Princeton Axion Search}
\author{Nate Otto}
\email{no7407@princeton.edu}
\affiliation{Department of Physics, Princeton University, Princeton, New Jersey 08544, USA}

\author{Jianyang Fu}
\affiliation{Department of Physics, Princeton University, Princeton, New Jersey 08544, USA}

\author{Joelle-Marie Begin}
\affiliation{Department of Physics, Princeton University, Princeton, New Jersey 08544, USA}

\author{Saptarshi Chaudhuri}
\email{sc25@princeton.edu}
\affiliation{Department of Physics, Princeton University, Princeton, New Jersey 08544, USA}

\author{Roman Kolevatov}
\affiliation{Department of Physics, Princeton University, Princeton, New Jersey 08544, USA}

\author{Lyman Page}
\affiliation{Department of Physics, Princeton University, Princeton, New Jersey 08544, USA}

\author{Joe Wiedemann}
\affiliation{Department of Physics, Princeton University, Princeton, New Jersey 08544, USA}

\date{\today}
\begin{abstract}
The Princeton Axion Search (PXS) targets the mass window \SIrange{0.8}{2.1}{\micro\electronvolt} (corresponding to frequencies \SIrange{200}{500}{\mega\hertz}) at DFSZ sensitivity. The experiment, a haloscope, will use a cylindrical copper-plated aluminum cavity of \SI{700}{\milli\meter} diameter and \SI{1022}{\milli\meter} height, immersed in a \SI{5}{\tesla} solenoidal magnetic field. A \SI{152}{\milli\meter} diameter tuning rod spanning the height of the cavity rotates to tune the primary mode resonance from \SIrange{390}{480}{\mega\hertz}, the frequency range for initial operations. We deduce the electric and magnetic field mode patterns using a combination of dielectric and metallic bead perturbations on the empty cavity. The measured mode frequencies and the $\text{TM}_{010}$ field profile, as a function of tuning rod position, are in excellent agreement with simulations. We describe the effect of fabrication and assembly imperfections, including the bowing of end plates and tilt of the tuning rod. The welded seams and end-cap joints are demonstrated to not significantly degrade $Q$. We combine measurement and simulation to assess an uncertainty on form factor across the full tuning range. Our results validate PXS's cavity architecture and inform future searches for sub-\si{\micro\electronvolt} axions, in which large-volume high-$Q$ structures are required to reach QCD sensitivity.
\end{abstract}

\maketitle

\section{Introduction}
\label{sec:Introduction}

Dark matter (DM) accounts for approximately \SI{85}{\percent} of the mass in the universe, yet its composition is unknown \citep{BODDY2022112}. The QCD axion, originally proposed by Peccei and Quinn as a solution to the strong CP problem, is a well motivated candidate  \cite{Peccei:1977hh,PhysRevLett.40.223,Wilczek:1977pj}.

Sikivie proposed the cavity haloscope method for direct detection of the axion in a lab setting via the Inverse Primakoff Effect \cite{PhysRevLett.51.1415}. In the presence of an external magnetic field, the conversion of axions to photons is enhanced when the resonant frequency of an embedded cavity matches the rest-mass frequency of the DM. For reference, a rest mass of \SI{1}{\micro\electronvolt} corresponds to \SI{240}{\mega\hertz}. The cavity's resonant frequency is tuned by changing its internal geometry, enabling the search for axions over a broad mass range \cite{doi:10.1126/sciadv.abm9928}.

The Princeton Axion Search (PXS) will probe rest-mass frequencies of \SIrange{200}{500}{\mega\hertz} at DFSZ sensitivity \citep{Dine:1981rt,Zhitnitsky:1980tq}, a community benchmark on the unknown axion-photon coupling\cite{ADMX:2023rgo,HAYSTAC:2018rwy,CAPP:2020utb,HAYSTAC:2020kwv,Adair:2022rtw}. Haloscopes aimed at similar and somewhat higher frequencies include ADMX \cite{chakrabarty2024low}, FLASH \cite{Alesini:2023qed}, and BabyIAXO \cite{Ahyoune:2023gfw}.

In collaboration with Princeton Plasma Physics Laboratory (PPPL), we are constructing a \SI{5}{\tesla}, \SI{740}{\milli\meter} bore solenoidal magnet. The large bore enables a \SI{700}{\milli\meter} diameter cavity, and facilitates the first PXS science scan covering \SIrange{390}{480}{\mega\hertz} (\SIrange{1.6}{2.0}{\micro\electronvolt} mass range). Future science runs will cover the remainder of the range. The cavity will be read out with state-of-the-art quantum-limited parametric amplifiers new to the frequency range\cite{faramarzi2025near}. 

The signal power for axion-photon conversion scales as \cite{sikivie2021invisible}
\begin{equation}
    P_{\mathrm{axion}}\propto  C^2 V^2 Q,
    \label{eq:Paxion}
\end{equation}
where $C$ is the cavity form factor, $V$ is the cavity mode volume, and $Q$ is the quality factor. After presenting the details of a simple and inexpensive cavity fabrication, the paper focuses on determining $Q$ and $C$ for our search. Section~\ref{sec:Instrument} describes the instrument --- the rolled-and-welded cavity and the tuning-rod mechanism. Section~\ref{sec:Performance} validates the empty-cavity performance and characterizes the mode spectrum, quality factor, and antenna coupling across the tuning range. Section~\ref{sec:FormFactor} presents the bead-pull characterization of both the empty cavity mode field profile, and with the tuning rod pivoted across the frequency range. We consider the effects of fabrication imperfections on field shape, mode crossings, and form factor. We conclude in Section~\ref{sec:Conclusions}.

\section{Instrument}
\label{sec:Instrument}

\subsection{Rolled-and-Welded Cavity}
\label{subsec:fabrication}
\begin{figure}
  \includegraphics[width=1\linewidth]{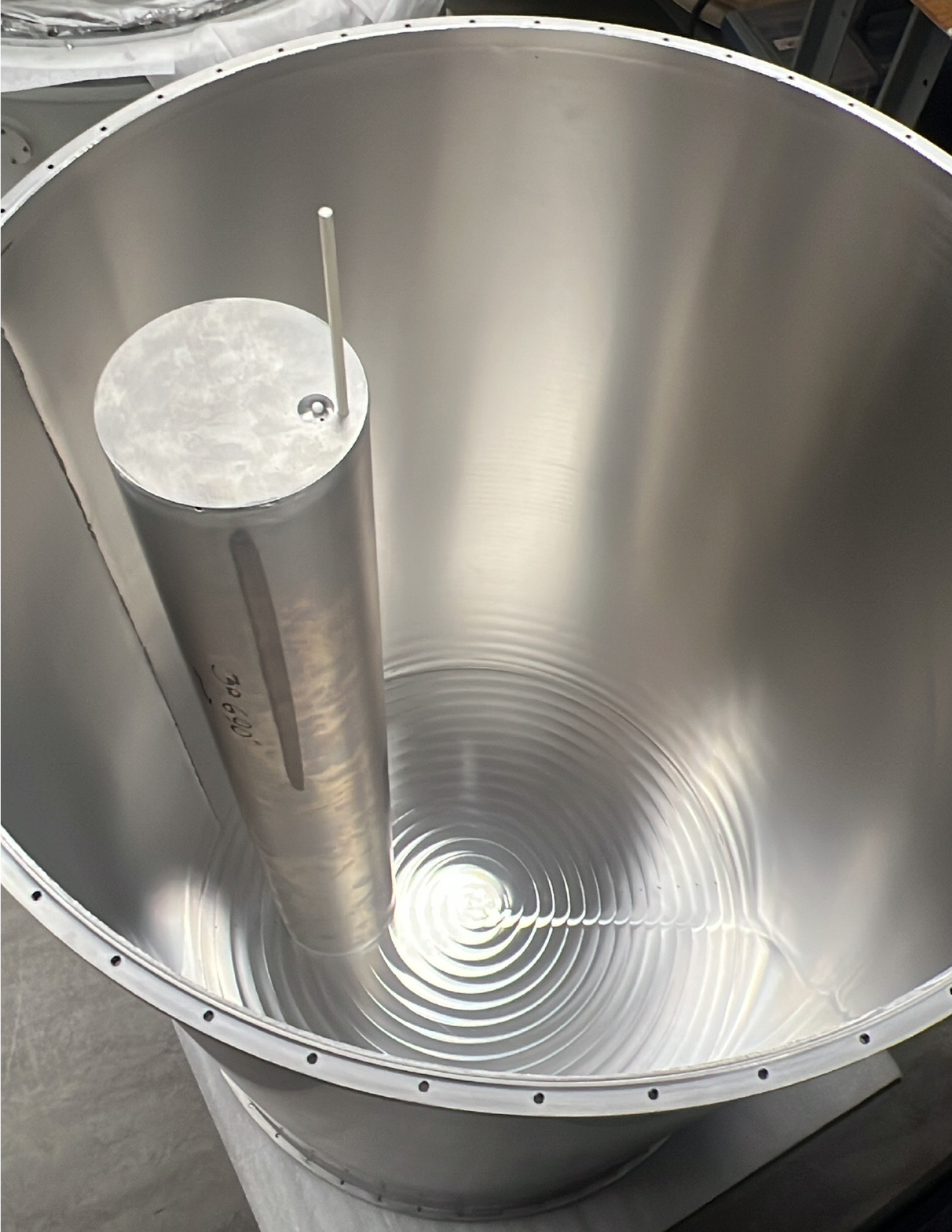}
  \caption{The PXS prototype cavity has \SI{1022}{\milli\meter} interior height and \SI{700}{\milli\meter} diameter. The top plate is not shown and the bottom end plate is attached to the cavity barrel by the stepped flange interface described in Fig.~\ref{fig:stepped flange}. The tuning rod rotates about the \SI{6.4}{\milli\meter} axle protruding from its end as seen in the figure.}
  \label{fig:cavity with tuning rod}
\end{figure}

\begin{figure}
  \includegraphics[width=.99\linewidth]{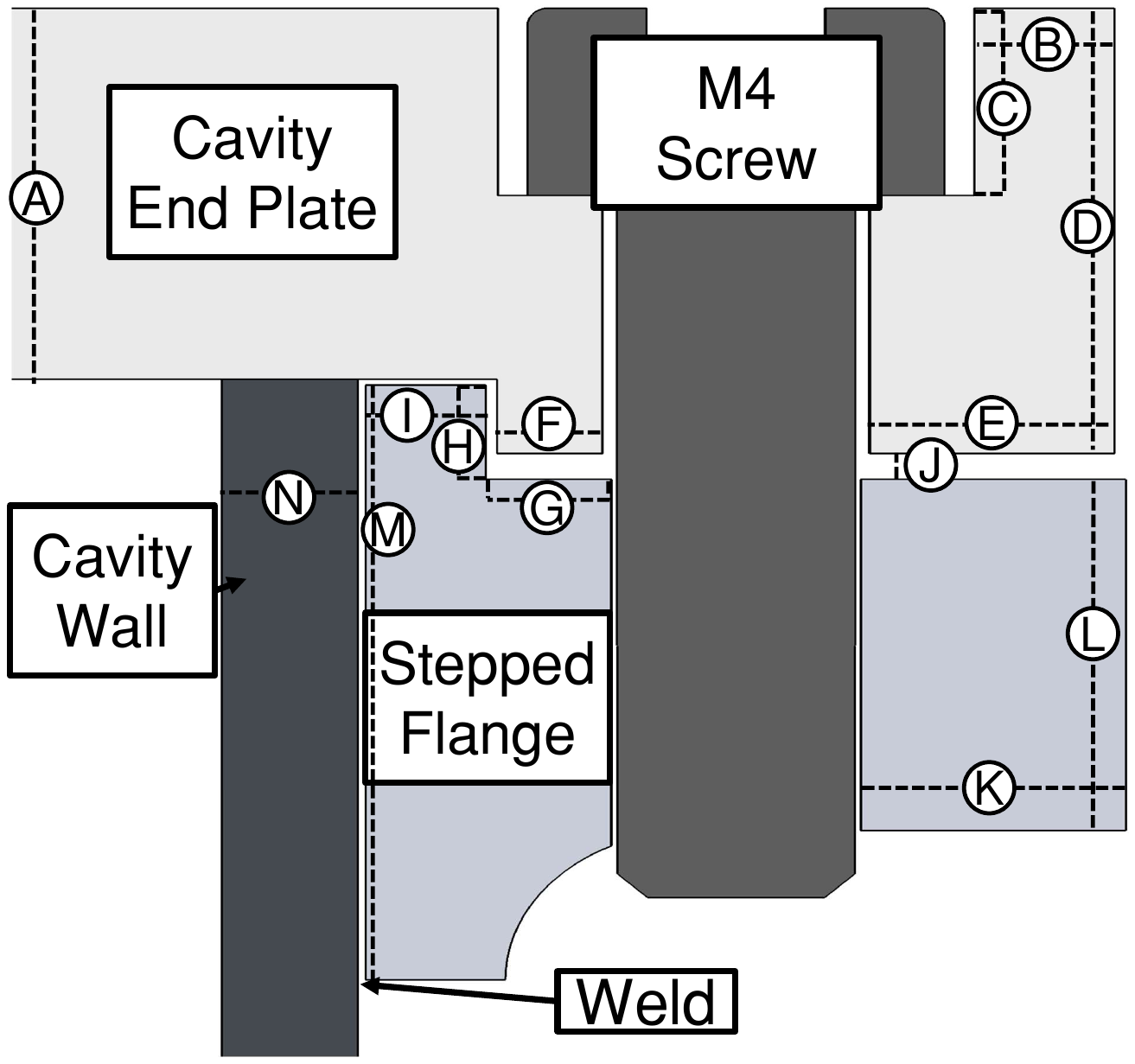}
  \caption{Schematic of the stepped flange interface between the cavity end plate and barrel. The stepped flange is welded onto the cavity barrel, and the lip has a height and width of \SI{2}{\milli\meter}. Labeled dimensions (mm) are given below.}
  \label{fig:stepped flange}
\centering
  \begin{tabular}{cccccccccccccc}
    \hline
    $\mathrm{A}$ & $\mathrm{B}$ & $\mathrm{C}$ & $\mathrm{D}$ & $\mathrm{E}$ & $\mathrm{F}$ & $\mathrm{G}$ & $\mathrm{H}$ & $\mathrm{I}$ & $\mathrm{J}$ & $\mathrm{K}$ & $\mathrm{L}$ & $\mathrm{M}$ & $\mathrm{N}$ \\
    \hline
    8.3 & 2.4 & 4.0 & 9.9 & 4.1 & 1.8 & 2.1 & 2.0 & 2.0 & 0.5 & 4.4 & 7.5 & 16 & 2.3 \\
    \hline
  \end{tabular}
\end{figure}

A typical fabrication for cavity haloscopes involves a seamless stainless steel barrel that is copper-plated (e.g., \cite{PhysRevD.42.1297}). To limit thermal noise and achieve DFSZ sensitivity, the PXS cavity needs to be cooled to temperatures below \SI{100}{\milli\kelvin}. To minimize thermal mass and limit costs, we use a rolled and welded Al-1100 barrel of \SI{2.3}{\milli\meter} thickness \footnote{Built by Precision Cryogenics (7804 Rockville Rd Indianapolis IN 46214).} that is then copper plated with at least \SI{10}{\micro\meter} thickness. The cavity is \SI{700}{\milli\meter} in internal diameter and \SI{1022}{\milli\meter} in height (see Fig. \ref{fig:cavity with tuning rod}). The end plates are \SI{8}{\milli\meter} thick Al-6061 and are bolted to the barrel through a stepped flange joint and 48 M4 screws (see Fig. \ref{fig:stepped flange}). The screws are torqued to \SI{2.3}{N.m}. The end plates bow measurably under this torque, making the cavity height non-constant: the nominal \SI{1020}{\milli\meter} height rises to roughly \SI{1022}{\milli\meter} at maximum end-plate deflection \footnote{The bowing would be reduced by adding a boss near ``E'' in Fig.~\ref{fig:stepped flange}.}. The volume of the empty cavity is \SI{393}{\liter}.

\subsection{The Tuning Rod}
\label{subsec:The Tuning Rod}

The cavity geometry is chosen to maximize Eq.\eqref{eq:Paxion}. PXS, like most other haloscopes, will tune the $\text{TM}_{010}$ resonant mode across the frequency range \cite{sikivie2021invisible}. The tuning rod is an Al-6061 cylinder with \SI{152}{\milli\meter} diameter and \SI{1015}{\milli\meter} tall. End caps of \SI{6.4}{\milli\meter} thickness are welded onto each end of a \SI{1.6}{\milli\meter} thick seamless tube. The tuning rod reduces the effective mode volume [Eq.~\ref{eq:Paxion}] modestly, from \SI{393}{\liter} to \SI{374}{\liter}. The rod height is \SI{7}{\milli\meter} less than the barrel height to allow controlled movement within the cavity without contact to the end plates. The mass of the tuning rod is \SI{2.5}{\kilo\gram} and the cavity \SI{35}{\kilo\gram}.

Through simulation, we found that this diameter of tuning rod, based on commercially available stock, can scan the first PXS science band of \SIrange{390}{480}{\mega\hertz} with a fixed pivot point. An axle of \SI{6.4}{\milli\meter} diameter, centered \SI{13}{\milli\meter} inward from the edge of the rod, couples the tuning rod to a precision motor outside the cavity; a flexible coupler outside the cryostat links the axle to the motor for cryogenic operation. G-10 and nylon axles have been used at room temperature to minimize radiation leakage, and a sapphire axle will be used cryogenically. A metallic tuning axle was found to increase radiation out of the cavity and substantially degrade the quality factor \cite{simanovskaia2021symmetric,hagmann1990cavity}. The tuning rod is electrically isolated from the cavity and will be cooled by contact to a retractable copper rod.

The pivot point is \SI{127}{\milli\meter} away from the center of the cavity. As the rod rotates, its center sweeps from \SI{64}{\milli\meter} to \SI{191}{\milli\meter} from the cavity center. We parametrize this displacement by the rotation angle: minimal displacement at $0^\circ$ and maximum displacement at $180^\circ$ (one half revolution). The top of the rod is measured to be \SI{4.7}{\milli\meter} below the top end plate, and the bottom rests on a sapphire ball with a \SI{2.4}{\milli\meter} clearance to the bottom plate. The sapphire bearing at the bottom prevents electrical conductive contact between the rod and the bottom end plate while providing mechanical and cooling support to the tuning rod.

The pivoted geometry introduces imperfections from an idealized cavity-rod system. It is known that vertical de-centering of the tuning rod position and tuning rod tilt within the cavity may degrade performance \cite{Rapidis:2018dci, Dyson:2024vhx}. The observed bowing, out-of-roundness, and weld seams may contribute additional degradation, motivating our study.

\section{Quality Factor and Mode Spectrum}
\label{sec:Performance}

\subsection{Quality Factor and Resonances}
\label{subsec:modeQ}

We present the quality factor and mode structure of the measurable cavity parameters for both the empty and tuned cavity. We decompose the loaded (total) quality factor $Q_L$ into internal and coupling contributions,
\begin{equation}
1/Q_L = 1/Q_c + 1/Q_i.
\label{eq:Q_decomposition}
\end{equation}
The coupling quality factor, $Q_c$, represents the loss associated with transmission from the antenna readout (see Sec.~\ref{subsec:res_antenna}). The internal quality factor, $Q_i$, is a quantity set by the cavity material conductivity, geometry, engagement of the interface between cavity end plates and the barrel, and radiated leakage from holes in the end plates (e.g., for the tuning rod).

Measurements for different cavity configurations are performed by connecting the cavity antenna to a vector network analyzer (VNA) \footnote{We use the Siglent SVA1032X in the Vector Network Analysis mode}. A wideband scan captures all coupled modes for a given configuration. Then, a fine trace taken around the resonance of each mode ($\text{TM}_{010}$ and otherwise) is fit to an In-phase and Quadrature (IQ) resonator model to extract the frequency, $Q_L$, $Q_c$, and $Q_i$ \cite{gao2008physics, probst2015efficient, casali2016characterization}. The recorded spectrum is corrected for cable resonances and standing-wave reflections through cable-response de-embedding based on a short-open-load (SOL) calibration, enabling a clearer view of the mode structure \cite{Probst:2014jga} \footnote{The signal for axion sensitivity does not require the SOL calibration, which is challenging to perform cryogenically.}.

In the empty resonator, surface losses on the aluminum barrel and the top and bottom end plates dominate. For the tuned cavity an additional rod-surface loss term enters, resulting in further reduction to $Q$.
\subsection{The Empty Cavity}
\label{subsec:res_empty}

We simulate the aluminum cavity in COMSOL Multiphysics\textsuperscript{\textregistered}, a package that solves Maxwell's equations by the finite element method (FEM) \cite{comsol}. Some computational details are given in the Appendix. The measured $Q_i$ agrees within \SI{10}{\percent} of both the analytic and simulated values, as summarized in Table~\ref{table:comparison}. Analytic and simulated values assume the cavity is a perfect cylinder with all losses conductive. The agreement demonstrates that the conductive losses dominate. A dedicated study of radiation leakage from the antenna hole measured $\lesssim \SI{2}{\percent}$ degradation on $Q_i$ for the empty cavity. Importantly, we find that the fabrication choices (rolled-and-welded cavity barrel, stepped flange joints) are sufficient for our target sensitivity.

\begin{table}[h!]
\centering
\renewcommand{\arraystretch}{2.5}
\resizebox{\columnwidth}{!}{%
\begin{tabular}{|c|c|c|c|}
\hline
& \textbf{Analytic} & \textbf{Simulated} & \textbf{Measured} \\
\hline
\textbf{Resonant Frequency (\si{\mega\hertz})} & \large{327.8} & \large{327.8} & \large{327.7} \\
\hline
\textbf{Quality Factor ($Q_i$)} & \large{52,600} & \large{52,500} & \large{48,000} \\
\hline
\textbf{Form Factor ($C$)} & \large{0.69} & \large{0.69} & \large{See Sec. \ref{sec:FormFactor}} \\
\hline
\end{tabular}
}
\caption{Comparison of analytic, simulated, and measured values of the empty aluminum cavity. The derived form factor values assume a uniform applied magnetic field. Valid field measurements in all of space within the cavity are required for a measurement of $C$.}
\label{table:comparison}
\end{table}

Uncertainty in the conductivity of the aluminum barrel and end plates can result in the discrepancy on the quality factor. The small difference in the measured frequency can arise from perturbations in cavity roundness and antenna insertion. 

\subsection{The Tuned Cavity}
\label{subsec:The Tuned Cavity}

The tuning rod angle defined in Sec.~\ref{subsec:The Tuning Rod} is stepped through one full revolution in less than $0.5^\circ$ increments to produce the mode map across the tuning range. A 70:1 gearbox reduction is attached to the tuning axle to further increase the angular resolution of the rod within the cavity. Mode crossings appear as resonances of similar frequencies to the $\text{TM}_{010}$ mode, but they have a different spatial pattern (TE, TEM). The interaction between the $\text{TM}_{010}$ mode and other modes limits the usable tuning range of each PXS geometry.

\subsubsection{Mode Map}

\begin{figure*}
\centering
  \includegraphics[width=0.99\linewidth]{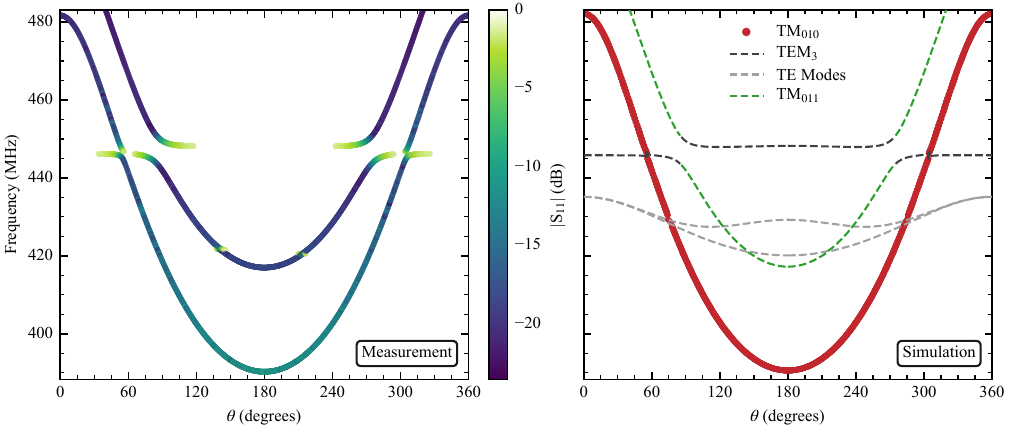}
  \caption{Measured (left) and simulated (right) mode maps over one full tuning rod revolution. The horizontal axis is the rod angle; \ang{0} corresponds to the rod nearest the cavity center, \ang{180} to the rod farthest from center. The tuning rod is stepped in $\approx \ang{0.1}$ increments for this measurement and $\ang{0.5}$ for simulation. Mode crossings with $\text{TEM}_{3}$ at $\SI{445}{\mega\hertz}$ ($\theta\!\approx\!\ang{55}$) and with TE modes at $\SI{428}{\mega\hertz}$ ($\theta\!\approx\!\ang{77}$) are visible. The two TE modes are weakly coupled to the vertical antenna and appear in measurement only near these crossings. Resonance widths are widened in the plot for visibility. The color-scale represents the maximum $S_{11}$ gain of each mode. The second half revolution retraces the same frequency range in reverse, providing an internal consistency check on the azimuthal symmetry of the cavity.}
  \label{fig:ModeMapSimMeas}
\end{figure*}

Figure~\ref{fig:ModeMapSimMeas} shows the measured mode map alongside the corresponding COMSOL simulation. We observe two mode crossings with the $\text{TM}_{010}$ mode: a $\text{TEM}_{3}$ resonance near \SI{445}{\mega\hertz} and a $\text{TE}$ resonance near \SI{428}{\mega\hertz}. We identify crossings by comparison with simulated frequency, the measured quality factor, and the use of bead pull described in Sec.~\ref{sec:FormFactor}.

\subsubsection{Quality Factor}
\begin{figure}

  \includegraphics[width=0.99\linewidth]{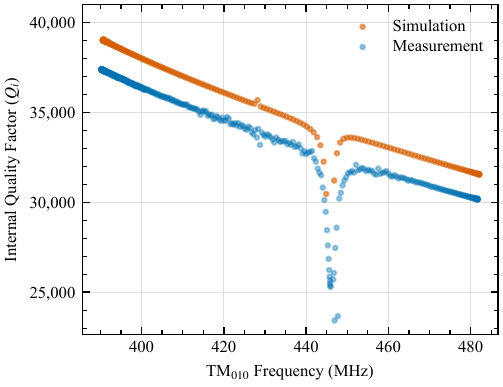}
  \caption{Measured internal quality factor (blue) of the $\text{TM}_{010}$ mode across the \SIrange{390}{480}{\mega\hertz} tuning range. The measurement values are at $\approx \ang{0.1}$ tuning rod increments and have an applied rolling average (\ang{0.5} window). Simulated $Q_i$ (orange) captures the trend of quality factor and effects from the mode crossing, but underestimates the width of the crossing around \SI{445}{\mega\hertz} (see Sec.~\ref{subsec: Mode Crossings} for more detail on mode crossings).}
  \label{fig:Measured Quality Factors}
\end{figure}

Figure~\ref{fig:Measured Quality Factors} shows $Q_i$ of the $\text{TM}_{010}$ mode as a function of mode frequency. Outside of mode crossings, the internal $Q$ is $\gtrsim 30,000$ and agrees with simulation at the \SI{10}{\percent} level.

\subsubsection{Antenna Coupling}
\label{subsec:res_antenna}

Cavity haloscopes have traditionally operated with one or more pickup antennae that are moved during the data-taking process \cite{ADMX:2025vom}. Achieving a desired coupling $\beta$, where $\beta = Q_i/Q_c$, often requires continuous cryogenic actuation of the antenna insertion depth, which adds operational overhead and reduces live time \cite{ADMX:2020ote}. Two coupling values are common for haloscope experiments: a critical coupling ($\beta = 1$), where the cavity presents a matched \SI{50}{\ohm} impedance on resonance, and an over-coupling ($\beta \approx 2$), which optimizes scan rate \cite{AlKenany:2016trt,chaudhuri2018optimal}.
\begin{figure}
  \includegraphics[width=0.99\linewidth]{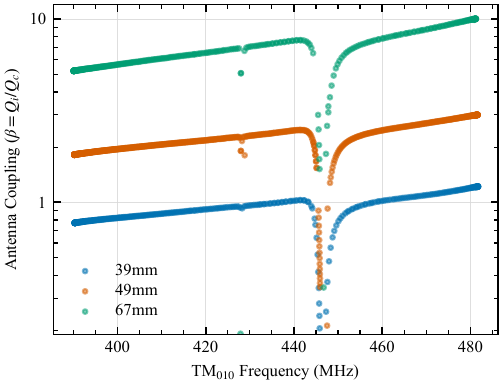}
  \caption{The measured antenna coupling, $\beta$, for different antenna lengths inside the cavity as a function of $\text{TM}_{010}$ frequency. Measurements are made at tuning rod steps of $\approx \ang{0.5}$.}
  \label{fig:Antenna Coupling}
\end{figure}

The antenna itself is a copper wire of diameter \SI{1.3}{\milli\meter}, soldered to the center pin of a standard SMA panel-mount jack (Amphenol RF 132146) and visible in Fig.~\ref{fig:bead pull setup}. We test various antenna lengths and their related couplings in Fig.~\ref{fig:Antenna Coupling}. The ability to maintain a relatively constant coupling across the full tuning range with a fixed antenna insertion depth mitigates the need for cryogenic motion and improves live time.

\section{Bead-Pull Field Characterization}
\label{sec:FormFactor}

We now consider the form factor, $C$, that describes the overlap between the electric field of the $\text{TM}_{010}$ mode and the DC magnetic field \cite{Stern:2015kzo},

\begin{equation}
  C = \frac{\big|\int_V \mathbf{E}(\mathbf{r})\!\cdot\!\mathbf{B}(\mathbf{r})\,dV\big|^2}{\big(\int_V |\mathbf{B}(\mathbf{r})|^2\,dV\big) \int_V \varepsilon_0 |\mathbf{E}|^2\,dV}.
  \label{eq:Cdef}
\end{equation}

This is difficult to measure directly in lab, owing to the vector nature of the fields. Haloscope searches thus rely on finite-element software to simulate the field profile and predict the resonant mode structure \cite{comsol,jones2020microwave,Schoeberl:2014ngsolve}. For the remainder of this paper, we assume a uniform external magnetic field, $\mathbf{B}(\mathbf{r}) \rightarrow \mathbf{B}_0\,\hat{z}$, across the length of the cavity. The PXS magnet's DC magnetic field will be used in Eq.~\eqref{eq:Cdef} for operations.

Efforts to probe the field structure of resonant cavities have a long history within accelerator physics, where bead perturbation provides a mapping of relative field strength \cite{maier1952field,doi:10.1049/pi-c.1960.0041,Du:2024sdb}. Similar tests have been applied to cavities of smaller volumes and higher frequencies \cite{rapidis2018application,Rapidis:2018dci,osti_2504569,garcíabarceló2026vortexcavityradesaxion,QUAX:2020wfd}. We adapt these techniques to probe mode structure for lower frequencies and larger volumes using two beads of different materials \cite{Nasserdine2014FieldMW}. We establish a procedure for separately probing electric and magnetic field profiles, and setting bounds on form factor in lab.

\subsection{Slater Perturbation Theory}
\label{subsec:slater_method}
Slater perturbation theory relates frequency shifts from small volume perturbations to field strength measurements \cite{RevModPhys.18.441}. We apply this technique to the cavity described in Sec.~\ref{sec:Instrument} for a homogeneous sphere of radius $a_b$ with relative permittivity $\varepsilon_r$ and permeability $\mu_r$. We retain only the real part of the Slater perturbation, the resonant-frequency shift; the imaginary part corresponds to a perturbation of the cavity losses and is negligible for the low-loss dielectric and metallic beads used here. Through the Clausius--Mossotti factors, $F_E=(\varepsilon_r-1)/(\varepsilon_r+2)$ and $F_H=(\mu_r-1)/(\mu_r+2)$ Slater perturbation theory reduces to:
\begin{equation}
\frac{\Delta f}{f_0}
\approx-\frac{\pi a_b^3}{U}
\bigl(\varepsilon_0 F_E\,|\mathbf{E}(\mathbf{r})|^{2}
     +\mu_0 F_H\,|\mathbf{H}(\mathbf{r})|^{2}\bigr).
\label{eq:sphere_shift}
\end{equation}
For the empty-cavity study (Fig.~\ref{fig:EH_disentangled}) we normalize the stored energy, U such that the relative field strengths have a maximum value of unity \footnote{Several haloscope bead-pull analyses \cite{simanovskaia2021symmetric, osti_2504569} instead adopt the dilute-limit form $\Delta f/f \propto (\varepsilon_r-1)/2$, which omits the $(\varepsilon_r+2)/3$ depolarization correction; for PTFE the dilute-limit prefactor is larger than the full Clausius--Mossotti value $F_E$ by a factor of $1.37$, so adopting it would overestimate $|\mathbf{E}|^2$ by the same factor --- a difference that matters at the precision targeted here.}. A dielectric polytetrafluoroethylene (PTFE) bead has $\mu_r = 1$ and probes $|\mathbf{E}|^{2}$ alone. In the metallic limit, $F_E\to 1$ and $F_H\to -1/2$, so an aluminum bead couples to both $|\mathbf{E}|^{2}$ and $|\mathbf{H}|^{2}$.
\subsubsection{Field Disentanglement}
\label{subsubsec:dual_bead}
We divide by the empty cavity frequency and write the fractional frequency shifts from the PTFE and metallic beads as a function of radial position in the cavity, as $S_D(r)$ and $S_M(r)$, respectively. The isolated field magnitudes as a function of the measured frequency shifts are
\begin{equation}
|\mathbf{E}(r)|^{2}
= -\frac{S_D(r)}{K F_E},
\qquad
|\mathbf{H}(r)|^{2}
= \frac{S_M(r) - S_D(r)/F_E}{K\alpha},
\label{eq:dual_bead_inversion}
\end{equation}
where $K \equiv \pi a_b^3\varepsilon_0/U$ and $\alpha\equiv\mu_0/(2\varepsilon_0)$ account for the geometric and material properties of the bead. The factor $F_E$ in Eq.~\eqref{eq:dual_bead_inversion} is that of the PTFE bead. 
This technique is demonstrated in Sec.~\ref{subsec:res_empty_beadpull}, where the analytic $\text{TM}_{010}$ fields of an empty cavity provide a direct comparison to measurement. The tuned-cavity analysis of Sec.~\ref{subsec:res_beadpull_tuning} then uses a PTFE bead to probe the $\text{TM}_{010}$ mode across the tuning range.

\subsection{Bead-Pull Apparatus}
\label{subsec:bp_apparatus}

The PTFE and aluminum beads of radius $a_b=\SI{32}{\milli\meter}$ are suspended on Kevlar string from a Teknic servo motor (CPM-SCSK-2310S-ELNA) with sub-millimeter level control. Each bead is inserted into the cavity with the top plate removed. The Kevlar string attached to the bead is then threaded through one of the holes on the top plate at a known radial position and attached to the servo motor, aligned with the hole and the bead. Holes of \SI{1.8}{\milli\meter} diameter are spaced at \SI{25}{\milli\meter} radial increments in the top and bottom end plates across the full cavity diameter. A total of 13 radial positions are used, from the cavity center out to \SI{300}{\milli\meter}.

\begin{figure}
  \includegraphics[width=0.99\linewidth]{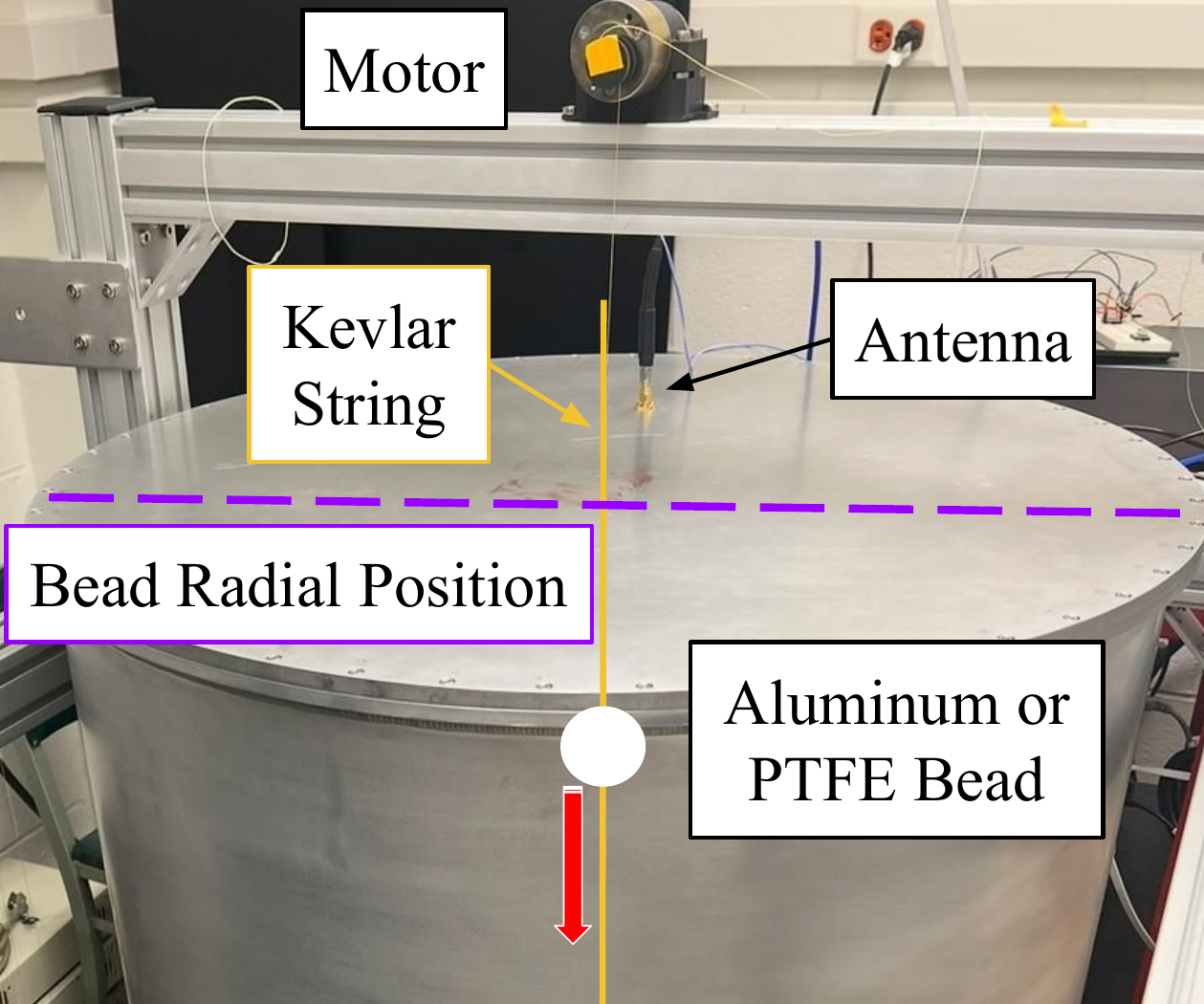}
  \caption{Labeled image of the bead-pull setup. The motor drives a PTFE or aluminum bead on a Kevlar string from resting on the bottom of the cavity to a few millimeters below the top plate at speeds $\leq$ \SI{7}{mm/s}. The same path and speed is retraced in the opposite direction. The VNA records the $\text{TM}_{010}$ resonant frequency as the bead traverses the cavity.}
  \label{fig:bead pull setup}
\end{figure}

The VNA records the $\text{TM}_{010}$ resonant frequency at roughly one measurement per second for each direction of traverse (see Fig.~\ref{fig:bead pull setup}). Each up-and-down traverse of the bead constitutes one run; at least five runs (and typically more than eight) are acquired per radial position per bead material. Measurements taken on the opposite side of the cavity (negative radial values) confirmed the symmetry of the $\text{TM}_{010}$ mode in the empty cavity. A data cut excludes all measurements made when a bead is within \SI{110}{\milli\meter} of each end plate, where Slater perturbation approximation breaks down \cite{sims2024dual,Karatay:2019yqs}. This same window is used in later sections. Applying the \SI{110}{\milli\meter} exclusion with bead proximity to the cavity barrel had no noticeable effect on the conclusions.

\subsubsection{Bead-Pull Field Mapping of the Empty Cavity}
\label{subsec:res_empty_beadpull}

The empty-cavity resonance is measured at $f_0=\SI{327.694}{\mega\hertz}$ at \SI{293}{\kelvin}. We monitor the cavity temperature during measurements. When the cavity heats by \SI{1}{\kelvin} and expands, the frequency drops by \SI{5}{\kilo\hertz} ($\lesssim \SI{.001}{\percent}$). We assume temperature-induced frequency drifts average out across runs.

\begin{figure}
  \includegraphics[width=0.99\linewidth]{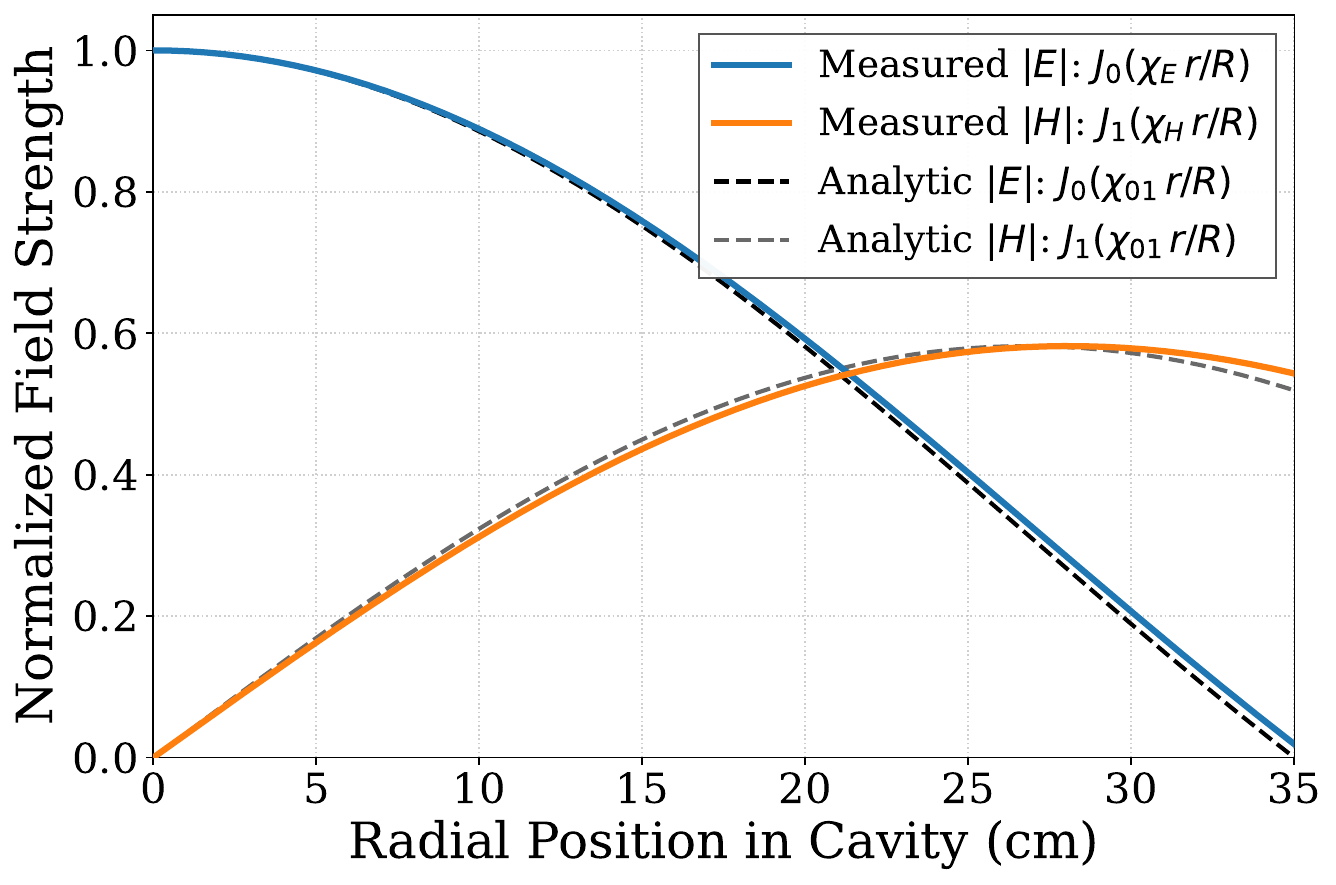}
  \caption{The normalized electric and magnetic field magnitudes of the analytic $\text{TM}_{010}$ (dashed, black) and the measured profiles from Eq.~\eqref{eq:dual_bead_inversion} (colored, solid) along the radial axis of the cavity. The first zero of the $J_0$ Bessel function, $\chi_{01}$ = 2.40 in comparison to the best fit values for the field structure ($\chi_E$, $\chi_H$) of 2.37. The cavity radius, R, is \SI{350}{\milli\meter}.}
  \label{fig:EH_disentangled}
\end{figure}

The fractional frequency shift, averaged over vertical positions within the analysis window and bead-pull runs, is applied to fit the analytic form of $\mathbf{E}$ and $\mathbf{H}$ for $\text{TM}_{010}$ (see Fig.~\ref{fig:EH_disentangled}) \footnote{The analytic solutions for the electric and magnetic field profiles of empty-cavity $\text{TM}_{010}$ are independent of vertical position within the cavity. The experimental measurements are consistent with this independence.}. Both fits yield $R^{2}\approx 0.999$ and are within the uncertainty of the dielectric constant of the PTFE bead ($\approx \SI{3}{\percent}$). The empty-cavity bead-pull thus establishes the validity of Slater perturbation theory in our apparatus. The technique to fully disentangle relative electric and magnetic field magnitudes using both bead types may be extended to other haloscopes and be of use in understanding mode crossings \cite{Kishimoto:2021ral}.

\subsection{Bead-Pulls Across the Tuning Range}
\label{subsec:res_beadpull_tuning}

\subsubsection{Bead-Pull for Tuned Cavity: Experimental Methods}

The same setup as \ref{subsec:bp_apparatus} is used, but with the tuning rod controlled by a separate Teknic motor (CPM-SCSK-2310S-EQNA). 
The bead is stepped vertically in \SI{30}{\milli\meter} intervals through the full cavity height; at each step we record the $\text{TM}_{010}$ frequency and quality factor. The tuning rod is then rotated by \ang{1} --- coarser than the $\lesssim \ang{0.1}$ steps of the mode map (Sec.~\ref{subsec:The Tuned Cavity}), since each bead-pull requires a full axial traverse --- and the process is repeated across the full tuning range. The cavity top plate is opened and closed along with a re-alignment of the tuning rod for 5 independent bead-pull runs. One run was taken with a deliberately applied rod tilt, so that the data ensemble samples realistic installation variability and a worst case alignment of the tuning rod.

Measurements presented use a bead radial location of \SI{200}{\milli\meter} from the center of the cavity (shown along the purple dashed line in Fig.~\ref{fig:bead pull setup}). The position is sufficiently far from the tuning rod that the measurements are valid within Slater perturbation theory while still in a high-field region across the full tuning range where the mode hybridization between $\text{TM}_{010}$ and TEM$_3$ is strongest. The internal quality factor is $\gtrsim 30,000$, confirming accurate resolution of the resonant frequency and a global retention of field structure \footnote{For the purposes of this paper we only consider changes to the resonant frequency in bead pull analysis, as opposed to also considering changes in quality factor.}.

\subsubsection{Bead-Pull for Tuned Cavity: Simulations}

COMSOL simulations provide the electric field throughout the cavity, and bead-pulls verify its magnitude along the measured path. Where mode crossings and fabrication imperfections are not fully captured by simulation, the measured and simulated profiles visibly disagree; quantifying this disagreement is the basis for the form-factor uncertainty developed below.

We use COMSOL to model the field profile for each tuning rod position. The resonant frequency of the $\text{TM}_{010}$ mode ($f_0(\theta)$), its electric field strength everywhere in space along the bead pull line ($E(\theta,z)$), total stored energy, ($U[f_0(\theta)]$), and form factor ($C(\theta)$) are then extracted at all tuning rod angles used in simulation. The predicted bead-pull frequency profile follows from Eq.~\eqref{eq:sphere_shift} as
\begin{equation}
\begin{aligned}
f_{\rm sim}(\theta, z) &= f_{0}(\theta) \left(1  -\frac{\pi\,{a_b}^3\,\varepsilon_{0}}{U[f_0(\theta)]}\,
|E(\theta, z)|^{2} \right).
\end{aligned}
\label{eq:fz}
\end{equation}
This $f_{\rm sim}(\theta,z)$ is the frequency shift expected from the perturbing PTFE bead and is compared with the frequency measurement of the bead-pull at the same location.

The measured and predicted $\text{TM}_{010}$ frequencies differ by an overall offset $\le\SI{1}{\mega\hertz}$; the median difference of \SI{200}{\kilo\hertz} is consistent with a sub-millimeter ($\lesssim$~\SI{0.3}{\milli\meter}) discrepancy between the fabricated cavity and the simulated \SI{700}{\milli\meter} diameter. Possible sources of error are presented in Table~\ref{table:systematics}.

We isolate this frequency offset through a single mean-alignment shift between the simulated ($f_{\rm sim}$) and measured ($f_{\rm data}$) bead-pull frequencies, $\Delta_{\rm offset} = \langle f_{\rm data}\rangle - \langle f_{\rm sim}\rangle$. This frequency offset is removed for the form-factor analysis \footnote{In Slater perturbation theory there is a referenced frequency, $f_0$, of the unperturbed mode. Upon insertion of the bead into the cavity, we lose knowledge of $f_0$ for the unperturbed frequency. The difference in overall frequency ($\Delta_{\text{offset}}$) applied to the formalism of the simulated shape (E- field) distribution is suppressed by 1/$f_0$ which removes the need to reconcile the value of $f_0$.}. We use the term ``aligned" to indicate we have applied $\Delta_{\text{offset}}$ to the simulation.

Eigenmode solutions on a finite tetrahedral mesh carry small-amplitude, non-physical numerical noise; within this window we linearly interpolate the simulated field-shift profile onto a common axial grid and apply $\Delta_{\text{offset}}$, suppressing mesh noise while preserving the mode shape.

\subsubsection{Initial Comparison}
\label{subsubsec: Initial Comparison}
\begin{figure}
  \includegraphics[width=0.99\linewidth]{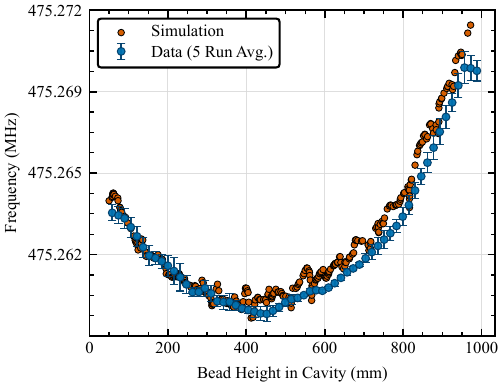}
  \caption{Comparison of measurements made with a PTFE bead-pull (blue) and simulation (red) at a rod angle of \ang{20}. The simulation assumes zero rod tilt and uses the expected cavity height. The frequency of the $\text{TM}_{010}$ mode ($f_0$) is \SI{475.3}{\mega\hertz}, well above the \SI{445}{\mega\hertz} mode crossing. Only considering the shape in mode across the range, the averaged data and aligned simulation agree within \SI{3.8}{\percent}  across the cavity height (see Eq.~\ref{eq:sigmacomp}). The frequency alignment, $\Delta_{\rm offset} =  \SI{310}{\kilo\hertz}$ and the form factor of the zero-rod-tilt simulation, $C_0$, is 0.543.}
  \label{fig:strong_agreement_above}
\end{figure}

At rod angles well above and below the TEM$_3$ mode crossing, the measured bead-pull profiles fall within \SI{4}{\percent} of simulation as shown in Fig.~\ref{fig:strong_agreement_above}. The simulated fields match data within the run-to-run noise, and the effects of tuning rod tilt on both field shape and form factor are negligible. When away from a mode crossing, we place an uncertainty on the form factor of $\lesssim \SI{4.1}{\percent}$, as described further in Sections \ref{subsec:form_factor_methods} and \ref{subsec:res_formfactor} below.

\subsection{Mode Crossings}
\label{subsec: Mode Crossings}

\begin{figure}
  \includegraphics[width=0.99\linewidth]{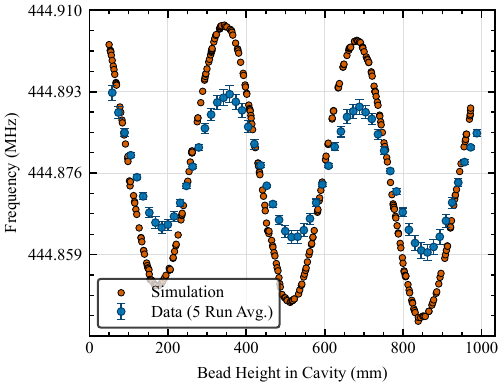}
  \caption{Comparison of measurements made with a PTFE bead-pull (blue) and simulation (red) at a rod angle of \ang{56}. The simulation assumes zero rod tilt and uses the expected cavity height, with $f_0 = \SI{445.0}{\mega\hertz}$. This rod angle is near the center of the mode crossing, and the measured profile shows clear mode-mixing structure. Data and aligned simulation agree within \SI{15}{\percent} across the cavity height; the simulation gives field strength greater than the measurement indicates. The frequency alignment, $\Delta_{\rm offset} = \SI{160}{\kilo\hertz}$, and the form factor of the zero-rod-tilt simulation, $C_0$, is 0.346.}
  \label{fig:Mode Crossing with Rod Tilt}
\end{figure}

As the tuning rod is moved toward the principal region with the $\text{TM}_{010}$--$\text{TEM}_{3}$ mode crossing, the measured $\text{TM}_{010}$ profile develops three axial nodes consistent with the higher-order mode mixture. This is evident in the structure of Fig.~\ref{fig:Mode Crossing with Rod Tilt}. From simulation we observe an enhanced sensitivity to rod positioning within the cavity around the mode crossing. The onset of a mode crossing is confirmed with the eigenmode simulations and measurements of Figs.~\ref{fig:ModeMapSimMeas}--\ref{fig:Antenna Coupling}.

At rod angles within the center of the mode crossing region, the \textit{pattern} of the field alignment between simulation and the hybridized mode is nonetheless excellent. This suggests that the solver captures the mode mixture where it is largest, or that the simulated geometry does not fully represent the physical cavity near the crossing. We therefore investigate this latter possibility.

\subsection{Form Factor Methods}
\label{subsec:form_factor_methods}
\subsubsection{Geometric Considerations}
\label{subsec:form_factor_method}

A typical concern for form-factor degradation and uncertainty considers tuning-rod tilt and the vertical location of the rod within the cavity \cite{Rapidis:2018dci,osti_2504569}. With the tuning rod seated on a sapphire ball bearing, the vertical position of the rod is constrained to within \SI{100}{\micro\meter}, an order of magnitude smaller than the finest standard mesh element size COMSOL uses for a geometry at this scale. Therefore, the variation in rod height and vertical position has no measurable effect on the simulated field profile.
There is variation in the internal cavity height up to \SI{2}{\milli\meter} when the end plates are torqued onto the cavity barrel. This is the largest geometric departure from our idealized cavity geometry. We account for this bowing with an adjusted cavity height that matches the measured separation distance between end plate and tuning rod. The cavity barrel is additionally measured to be out-of-round at the \SI{1}{\milli\meter} level; while this radial imperfection propagates into the overall $C^2 V^2Q $ sensitivity budget, it does not appreciably affect the bead-pull field comparisons.

\begin{table}[h!]
\centering
\renewcommand{\arraystretch}{1.5}
\begin{tabular}{|l|c|c|c|}
\hline
\textbf{Systematic} & \textbf{Variation} & \textbf{$f_0$} & \textbf{$\Delta f$} \\
\hline
Cavity Temperature           & \SI{1}{\kelvin}            & \SI{5}{\kilo\hertz}   & --          \\
\hline
PTFE Dielectric Constant     & 0.05                  & --            & \SI{1}{\kilo\hertz} \\
\hline
New Run Cavity Alignment            & 1                     & \SI{100}{\kilo\hertz} & \SI{1}{\kilo\hertz} \\
\hline
Measured Tuning Rod Position          & \ang{1}     & \SI{500}{\kilo\hertz} & \SI{3}{\kilo\hertz} \\
\hline
Simulated Tuning Rod Position & \ang{1}   & \SI{500}{\kilo\hertz} & \SI{3}{\kilo\hertz} \\
\hline
Measured Tuning Rod Tilt              & \ang{0.025}   & \SI{10}{\kilo\hertz}  & \SI{2}{\kilo\hertz} \\
\hline
Simulated Tuning Rod Tilt              & \ang{0.025}   & \SI{10}{\kilo\hertz}  & \SI{2}{\kilo\hertz} \\
\hline
Measured Bead Vertical Position  & \SI{30}{\milli\meter}            & --            & \SI{1}{\kilo\hertz} \\
\hline
Simulated Field Vertical Position  & \SI{10}{\milli\meter}            & --            & \SI{1}{\kilo\hertz} \\
\hline
Simulated Cavity Height & \SI{1}{\milli\meter}   & \SI{200}{\kilo\hertz} & \SI{3}{\kilo\hertz} \\
\hline
\end{tabular}
\caption{Physical and simulated deviations of parameters that contribute to changes in the measured $\text{TM}_{010}$ frequency from any bead-pull run. The listed values are typical fluctuations between bead-pull runs while varying one parameter and holding all others fixed. The column $f_0$ denotes a shift in the absolute resonant frequency; $\Delta f$ denotes the effect on the bead-pull frequency shift relative to $f_0$. }
\label{table:systematics}
\end{table}

A \SI{2}{\milli\meter} taller cavity alters the $\text{TEM}_3$ frequency across the tuning range by $\lesssim$ \SI{1}{\mega\hertz}; however, the width of the mode crossing increases by $\gtrsim$ \SI{5}{\mega\hertz}. We account for the cavity-height ambiguity through a dedicated height sweep (Sec.~\ref{subsubsec:form factor statistics}) and derive an uncertainty that is a component of the net uncertainty on $C(\theta)$ \cite{PhysRevD.106.092007}.

To quantify the effects of rod-tilt, we perform a dedicated simulation sweep of rod-tilt angles $(\phi_x, \phi_y)$ on a $5\times5$ grid spanning $\pm 0.05^\circ$ in steps of $0.025^\circ$ in each direction, at every rod angle in the tuning sweep. The mechanical tolerance on the rod axis is $\lesssim$ $0.06^\circ$, which corresponds to a horizontal misalignment of \SI{1}{\milli\meter}. This is the maximum displacement allowed by the hole through which the tuning axle enters the cavity. At each rod angle we extract the simulated form factor $C_i(\theta)$ for each tilt cell $i$, and take the spread of these values across the grid as the tilt-induced form-factor uncertainty. We remain agnostic to the tilt value that best matches the data, as the tilt is likely to change each time the cavity is opened and closed.

\subsubsection{Form Factor Uncertainty Metric}
\label{subsubsec:form factor statistics}
A two-component treatment is applied here --- (i) a simulation-based central value for the form factor and (ii) a combined uncertainty that absorbs the simulation and data mismatch from bead-pull measurements, the tuning rod tilt, and the cavity-height spread. We implement the variations and uncertainties from Table~\ref{table:systematics}.

Away from a mode crossing, our results are similar to refs.~\cite{ADMX:2023rgo,Adair:2022rtw,Bae:2024kmy,Bae:2025bqr,HAYSTAC:2025nrt}. We increase our uncertainty substantially when mode profile is uncertain.

We take the zero-rod-tilt, measured cavity-height simulation as the central form factor $C_0(\theta)$, and combine the three uncertainty sources in quadrature.

First, the bead-pull shape mismatch quantifies how well the measured field profile is reproduced by the no-tilt simulation. The measured ($f_{\rm data}$) and simulated ($f_{\rm sim}$) bead-pull frequencies are each linearly interpolated onto a common axial grid and ``aligned" to determine the fractional shape mismatch at rod angle $\theta$,
\begin{equation}
g(\theta) =
\frac{\bigl\langle\,\lvert f_{\rm data}(\theta,z) - f_{\rm sim}(\theta,z)\rvert\,\bigr\rangle_{[a,b]}}
     {\lvert \Delta f_{\rm dip}(\theta)\rvert} ,
\label{eq:sigmacomp}
\end{equation}
where $\langle\cdot\rangle_{[a,b]}$ denotes the mean over the analysis window and $\Delta f_{\rm dip}(\theta) = \max_{z} f_{\rm sim} - \min_{z} f_{\rm sim}$ is the full frequency swing of the simulated bead-pull profile over the analysis window, independent of the axial location of the extrema. For example, in Fig.~\ref{fig:strong_agreement_above}, $\Delta_{\rm offset} = \SI{310}{\kilo\hertz}$ and $\Delta f_{\rm dip} = \SI{12}{\kilo\hertz}$; in Fig.~\ref{fig:Mode Crossing with Rod Tilt} $\Delta_{\rm offset} = \SI{160}{\kilo\hertz}$ and $\Delta f_{\rm dip} = \SI{63}{\kilo\hertz}$. Thus, $g(\theta)$ measures the residual relative to the size of the perturbation signal. We convert this fractional quantity to form-factor units by scaling with the zero-rod-tilt, expected-cavity-height value, yielding $\sigma_{\rm bp}(\theta) = g(\theta)\, C_0(\theta)$.

Second, the rod-tilt uncertainty $\sigma_{\rm tilt}(\theta)$ is the standard deviation of the simulated form factor $C_i(\theta)$ across the tuning rod tilt cells, at each rod angle. Third, the cavity-height uncertainty $\sigma_{\rm height}(\theta)$ is the standard deviation of the form factor over a dedicated sweep of internal cavity heights, bracketing the measured cavity height (\SI{1022}{\milli\meter}) by $\pm 1$~mm.

The total form-factor uncertainty is the quadrature sum of the three sources,
\begin{equation}
\sigma_{\rm total}(\theta) =
\sqrt{\,\sigma_{\rm bp}^2(\theta) + \sigma_{\rm tilt}^2(\theta) + \sigma_{\rm height}^2(\theta)\,},
\label{eq:sigmatotal}
\end{equation}
and we quote the fractional form-factor uncertainty as $\sigma_{\rm total}(\theta)/C_0(\theta)$. Fig.~\ref{fig:C vs f0} shows the values of $C_0$ and $\sigma_{\rm total}$ as a function of the $\text{TM}_{010}$ frequency.

\subsection{Form Factor Analysis}
\label{subsec:res_formfactor}

\begin{figure}
\includegraphics[width=0.99\linewidth]{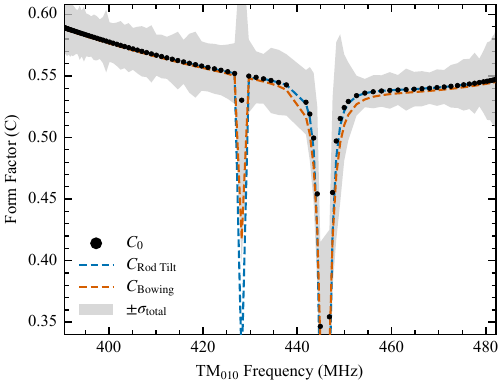}
\caption{The $\text{TM}_{010}$ form factor $C$ as a function of frequency across the full tuning range of the cavity. The blue points show the zero-rod-tilt, central-cavity-height form factor $C_0(\theta)$, plotted against the corresponding $\text{TM}_{010}$ frequency. The dashed blue and orange curves show the minimum form factor at each rod angle over the rod-tilt ensemble and the cavity-height sweep ($1022\pm1$~mm), respectively; they are discontinuous across the mode-crossing regions, where a single mode is not defined. The gray shaded band shows the total form factor uncertainty $\pm\sigma_{\rm total}$ from Eq.~\eqref{eq:sigmatotal}, the quadrature sum of the bead-pull shape mismatch, rod-tilt, and cavity-height contributions. Away from mode crossings, the uncertainties on $\sigma_{\rm tilt}$ and $\sigma_{\rm height}$ are negligible. The uncertainty increases near the two mode crossings (\SI{445}{\mega\hertz} and \SI{428}{\mega\hertz}), where we lose constraining power on $C$.}
\label{fig:C vs f0}
\end{figure}

Based on our measurements and simulations, the largest uncertainty on the form factor is the separation distance between the tuning rod and cavity end plates. Knowledge of this effect is based on a combination of simulation and bead-pulls. For example, Fig.~\ref{fig:Mode Crossing with Rod Tilt} shows the difference between simulation and reality. Applying the prescription of Eqs.~\eqref{eq:sigmacomp}--\eqref{eq:sigmatotal} across the full tuning range yields the form-factor envelope shown in Fig.~\ref{fig:C vs f0}.

Away from the mode crossings, all three uncertainty contributions are small: the bead-pull shape mismatch is $\lesssim \SI{4.0}{\percent}$, the rod-tilt spread is $\lesssim \SI{0.05}{\percent}$, and the cavity-height spread is $\lesssim \SI{0.4}{\percent}$, giving a total fractional uncertainty on $C$ of $\lesssim \SI{4.1}{\percent}$ (visible in Fig.~\ref{fig:strong_agreement_above}). Near the wider mode crossing ($\text{TM}_{010}$--$\text{TEM}_3$), the bead-pull mismatch and geometric spreads grow, and the total uncertainty rises to a maximum of $\SI{25}{\percent}$. The narrower crossing with the lowest TE cavity modes in the first PXS band occurs near \SI{428}{\mega\hertz}. At this mode crossing, the simulation accurately captures the mode pattern and the uncertainty is dominated by the tilt and cavity-height spreads.

As a separate per-angle diagnostic that does not enter the form-factor bound, the interpolated field profile of each individual tilt cell was evaluated against the bead-pull data and the best-matching tilt $(\phi_x, \phi_y)$ recorded. A pair recurring as the best match across many rod angles would be evidence of a persistent geometric offset of the physical rod from the nominal axis; the observed scatter is consistent with $(0,0)$ within the combined measurement and simulation uncertainty, justifying the use of the no-tilt geometry as the baseline $C_0(\theta)$.
\section{Conclusions}
\label{sec:Conclusions}
We have presented an end-to-end experimental and computational characterization of a rolled-and-welded aluminum cavity haloscope and its tuning mechanism for the Princeton Axion Search (PXS). Empty-cavity studies validated the frequency and field profiles of the $\text{TM}_{010}$ mode. With a pivoted \SI{152}{\milli\meter}-diameter aluminum tuning rod, this mode is tuned from \SIrange{390}{480}{\mega\hertz} with a measured $Q_{i}\approx 30,000$, in good agreement with simulations. A dense mode map over the full rod revolution reproduces all expected mode crossings.

Outside mode crossings, the simulated mode shape matches data, supporting the use of the simulated form factor with an added uncertainty guided by bead-pull measurements. Inside the principal $\text{TM}_{010}$--$\text{TEM}_{3}$ crossing, simulated and measured mode mismatch increases sharply, evidencing mode hybridization that the simulated geometry does not entirely capture. The presence of such mode crossings does not significantly disrupt the achievable science across the first PXS science target.

Techniques established here certify form factor and mode profile at this scale, demonstrating axion haloscopes and tunable resonators are capable of targeting lower masses with relaxed fabrication tolerances \cite{DMRadio:2022pkf,Ouellet:2018beu}.  

\section{Acknowledgments}
\label{sec:Acknowledgments}
We would like to thank the Princeton Machine Shop, Glenn Atkinson and Matthew Komor, and Pat Bradshaw and Steve Lowe with the student machine shop. We would also like to thank Andrew Liu and Sam Day-Weiss for helpful discussions throughout the writing of this paper. 
This work was supported by the Small Scale Experiments for Fundamental Physics initiative, jointly sponsored by the Gordon and Betty Moore, Simons, John Templeton, and Alfred P. Sloan Foundations. This project was funded by the Simons Foundation under grant number MP-TMPS-00002996. Additional support was provided by the Princeton Innovation Fund for New Ideas in the Natural Sciences. 
\section{Appendix}
\label{sec:Appendix}

\subsection{Finite Element Method}

The maximum element size of COMSOL's ``Extra Fine Mesh" at this scale of cavity and tuning rod is \SI{35.7}{\milli\meter}, and the finest default mesh choice "Extremely Fine Mesh" is \SI{20}{\milli\meter}. For detailed simulations across the tuning range, simulations using the Extremely Fine Mesh often crashed, as the computer either ran out of memory, or some other internal COMSOL error. These simulations are computationally expensive but accessible on high-end consumer-grade hardware. Our system was built around a i9-14900K CPU with \SI{128}{\giga\byte} of \SI{5400}{MT/s} DDR5 RAM. Our simulations typically occupied \SIrange{60}{90}{\giga\byte} of RAM during the solve. Each simulation results in a COMSOL file around \SI{150}{\giga\byte}, so we use cheaper HDD storage to have the capacity for numerous tests. Typical simulations take \SIrange{36}{48}{\hour}, and solve times are strongly reduced by high core-count CPUs. Exporting the solution is also time-consuming: the finer mesh requires additional interpolation, and writing the \SI{150}{\giga\byte} solution file to disk dominates the export time.

NGSolve, an open-source Python package, provides orders-of-magnitude faster studies, with comparable levels of output \cite{Schoeberl:2014ngsolve,Schoeberl1997Netgen}. However, there are differences in the solved frequencies of the $\text{TM}_{010}$ mode, with disagreements up to \SI{2}{\mega\hertz} ($\lesssim \SI{0.5}{\percent}$), so we rely on COMSOL for predicted quality factor, field strength, and form factor. For the overall form factor away from mode crossings, the two simulation tools differ by no more than \SI{0.3}{\percent}. At the mode crossing, there is an overall offset on the frequency of the crossing, which will be the subject of future investigations.

\bibliographystyle{apsrev4-2}
\bibliography{Bead_Pull_Paper}
\end{document}